# Electrical magnetochiral effect induced by chiral spin fluctuations

T. Yokouchi[1†], N. Kanazawa[1], A. Kikkawa[2], D. Morikawa[2], K. Shibata[2], T. Arima[2,3], Y. Taguchi[2], F. Kagawa[2], Y. Tokura[1,2†]

[1]*Department of Applied Physics, The University of Tokyo, Tokyo 113-8656, Japan.*

[2]*RIKEN Center for Emergent Matter Science (CEMS), Wako 351-0198, Japan.*

[3]*Department of Advanced Materials Science, The University of Tokyo, Kashiwa 277-8561, Japan*

† To whom correspondence should be addressed. E-mail: yokouchi@cmr.t.u-tokyo.ac.jp and tokura@riken.jp

**Chirality of matter can produce unique responses in optics[1], electricity[2] and magnetism[3]. In particular, magnetic crystals transmit their handedness to the magnetism via antisymmetric exchange interaction of relativistic origin, producing helical spin orders as well as their fluctuations. Here we report for a chiral magnet MnSi that chiral spin fluctuations manifest themselves in the electrical magnetochiral effect (eMChE)[4-5], *i.e.* the nonreciprocal and nonlinear response characterized by the electrical conductance depending on inner product of electric and magnetic fields *E*·*B*. Prominent eMChE signals emerge at specific temperature-magnetic field-pressure regions: in the paramagnetic phase just above the helical ordering temperature and in the partially-ordered topological spin state at low temperatures and high pressures, where thermal and quantum spin fluctuations are conspicuous in proximity of classical and quantum phase transitions, respectively. The finding of the asymmetric electron scattering by chiral spin fluctuations may explore new electromagnetic functionality in chiral magnets.**

Both spin fluctuations and chiral magnetism form fundamental concepts in condensed matter physics, being the sources of various significant physical phenomena. For example, antiferromagnetic spin fluctuations are involved in formation of Cooper pairs in the high-temperature superconducting cuprates[6], and quantum spin fluctuations break down the Fermi-liquid behaviour[7]. As for the chiral magnetism, real-space Berry phase related to non-coplanar spin textures with finite scalar spin chirality $\chi_{ijk} = \boldsymbol{S}_i \cdot (\boldsymbol{S}_j \times \boldsymbol{S}_k)$, where $\boldsymbol{S}_n$ $(n = i,\ j,\ k)$ are adjacent three spins, can produce emergent magnetic field and hence the topological Hall effect[8,9]. Furthermore, nonlinear large magnetoelectric effects are linked to chirality of spin configurations, *i.e.* vector spin chirality $\boldsymbol{C}_{ij} = \boldsymbol{S}_i \times \boldsymbol{S}_j$ (Ref. 2). Despite appreciation of these two concepts, cooperative phenomena from spin fluctuations and spin chirality have not fully been explored in charge transport phenomena. Here we demonstrate that thermal and quantum spin fluctuations endowed with finite vector spin chirality produce a directional nonlinear magnetotransport with the conductance proportional to inner product of magnetic field ($\boldsymbol{B}$) and electric field ($\boldsymbol{E}$), termed electrical magnetochiral effect (eMChE).

Spin structures and their dynamics in chiral-lattice magnets bear chiral nature due to antisymmetric exchange interactions, such as Dzyaloshinskii-Moriya (DM) interaction ($\boldsymbol{D} \cdot \boldsymbol{C}_{ij}$); the sign of the DM vector $\boldsymbol{D}$ is intrinsically dependent on the crystalline chirality. As a consequence, the sign of their magnetic chirality, as defined for example by $\boldsymbol{r}_{ij} \cdot \boldsymbol{C}_{ij}$ ($\boldsymbol{r}_{ij}$ being the vector connecting *i*-th and *j*-th sites), is macroscopically coherent throughout the crystal, which can make the chirality dependent transport signals macroscopically visible. MnSi of the present focus has the noncentrosymmetric lattice structure, which can exist in two enantiomeric forms: right- and left-handed structures as shown in Fig. 1**a**. Due to the competition between the ferromagnetic exchange interaction and the DM interaction, there emerge various spin winding structures, whose modulation directions, *i.e.* magnetic helicity, are determined by handedness of the corresponding lattice structures. Below the magnetic ordering temperature $T_c$ = 29.5 K, the long-period (~18 nm) helical spin structure (Fig. 1**b**) forms[10]. In addition, topological spin objects, skyrmions (Fig. 1**c**), condense in triangular-lattice (skyrmion-lattice state) at $0.1\ \mathrm{T} \lesssim B \lesssim 0.3\ \mathrm{T}$ just below $T_c$ (Ref. 11). Above $T_c$, where the long-range magnetic orders disappear, short-range spin correlations still survive without losing the chiral nature[12–15], as described as $\langle (\langle \boldsymbol{C}_{ij} \rangle - \boldsymbol{C}_{ij})^2 \rangle$. Strong enhancement of the chiral spin fluctuations around $T_c$ has been theoretically proposed[16] and demonstrated by polarized neutron scattering experiments[12–15].

We found that the chiral spin fluctuations play a key role in the eMChE in MnSi. From the viewpoint of symmetry, eMChE can generally appear in chiral systems. Resistivities with current density $\boldsymbol{j}$ parallel and antiparallel to magnetic field $\boldsymbol{B}$ exhibit different values[4, 5]. Resistivity considering eMChE can be described as follows:

$$\boldsymbol{E} = \rho[1 + \hat{\gamma}^{R/L}(B)(\boldsymbol{j} \cdot \boldsymbol{B})]\boldsymbol{j}. \qquad (1)$$

Here, $\rho$ is the linear term of longitudinal resistivity, *R* and *L* denote right- and left-handed crystalline chiralities, and $\hat{\gamma}^{R/L}$ (*B*) is the eMChE coefficient being an even function of *B*. We schematize the current-directional response in MnSi for each experimental configuration in Fig. 1**e**. Note that Eq. (1) can be transformed to the equivalent form, $\boldsymbol{j} = \left(\frac{1}{\rho}\right)[1 - \hat{\gamma}^{R/L}(B)(\boldsymbol{E} \cdot \boldsymbol{B})/\rho^2]\boldsymbol{E}$. Under time-reversal operation, the current direction for higher conductance is reversed. Likewise, the higher-conductance direction is opposite for different crystal chiralities; $\hat{\gamma}^{R}(B) = -\hat{\gamma}^{L}(B)$. Since voltage signals from eMChE are anticipated to be small, enough large current density is required to detect eMChE. In order to increase current density under the limitation of external high-precision current sources, by using focused ion beam (FIB) we fabricated microscale thin plates of MnSi, whose thickness and width are approximately 500 nm and 10 μm, respectively (Fig. 1**d**). Temperature dependence of resistivities and magnetic phase diagram of thin plate samples resemble those of bulk samples (see Supplementary Information),

First, we show typical profiles of eMChE signals observed in MnSi. Since eMChE appears as a nonlinear transport response in proportion to $j^2$ (Eq. 1), we measured second harmonic resistivity ($\rho^{2f}$), which is directly connected to eMChE as $\rho^{2f} = \frac{\rho}{2}\hat{\gamma}^{R/L}(B)(\boldsymbol{j} \cdot \boldsymbol{B})$ (see Supplementary Information). The magnetic field and current were applied parallel to [100] direction unless otherwise noted. Figures 2**a** and **b** present $\rho^{2f}$ of right- and left-handed MnSi at *T* = 35 K with current density $j = 1.0 \times 10^9$ A/m$^2$ and frequency *f* = 13 Hz. The both right- and left-handed crystals were selected from several batches by identifying the handedness in terms of the conversion beam electron diffraction method (see Supplementary Information). In accord with the expected contributions from eMChE, both the field profiles of $\rho^{2f}$ of the right- and left-handed crystals are antisymmetric against $\boldsymbol{B}$, exhibiting the opposite sign to each other. To further confirm that the observed $\rho^{2f}$ signals stem from eMChE, we measured $\rho^{2f}$ as functions of current density and relative angle $\theta$ between $\boldsymbol{B}$ and $\boldsymbol{j}$ both lying in-plane (Figs. 2**c** and **d**). Both the $j$- and $\theta$-dependences obey the expected behaviours from the relation $\rho^{2f} = \frac{\rho}{2}\hat{\gamma}^{R/L}(B)(\boldsymbol{j}\cdot\boldsymbol{B})$; $\rho^{2f}$ is proportional to $j$ and $\cos\theta$, respectively.

Next, we discuss a dominant mechanism of eMChE in MnSi. One mechanism proposed for eMChE in non-magnetic materials is so-called self-field effect[4]. In this mechanism, eMChE is expected to show $B$-linear dependence. This is however inconsistent with the present observation that $\rho^{2f}$ is suppressed at high magnetic field as presented in Figs. 2**a** and **b**. Another possible mechanism of eMChE is asymmetric electron scatterings by chiral scatterers[4]. To examine this, we investigate $T$- and $B$-dependences of $\rho^{2f}$. In Fig. 3**a**, we show a contour mapping of $\rho^{2f}$ in the $T$-$B$ plane for left-handed MnSi, measured with $j = 7.5\times10^{8}$ A/m$^2$. Second harmonic resistivity becomes prominent in the paramagnetic region, showing the broad peak profile in the $T$-$B$ plane just above the phase boundary (helical-to-paramagnetic) and the crossover line (ferromagnetic-to-paramagnetic). In contrast, the signal suddenly declines with entering the long-range ordering phases. These behaviours are exemplified by the $T$-scan of $\rho^{2f}$ at $B$ = 0.4 T as shown in Fig. 3**b**; the magnitude of $\rho^{2f}$ exhibits its maximum near $T_c$, and shows sharper decrease at the side of helical phase than at the side of paramagnetic phase. The above results indicate that eMChE in MnSi is related to the strongly enhanced chiral spin fluctuations around and immediately above $T_c$[12−15], which should induce asymmetric electron scatterings. It is worth noting here that eMChE also observed at the phase boundary between the conical and skyrmion-lattice states (see Fig. S3 in Supplementary Information).

Up to this point, we have revealed that the eMChE in MnSi arises from thermal spin fluctuations enhanced in the vicinity of the helical order as well as of the skyrmion-lattice phase. Next, we investigate the possible effect of quantum spin fluctuations on eMChE. In bulk samples of MnSi, the long-range static helical order is suppressed under pressure, and disappears at a pressure of $p$ = 14.6 kbar[17−19], where the quantum phase transition occurs and consequently the quantum spin fluctuations become dominant. Even above the pressure for this quantum phase transition, there exists a dynamical topological magnetic order, which fluctuates on time scales between $10^{-10}$ s and $10^{-11}$ s[17−19]. Since this dynamical magnetic order, called partial order (PO), is promoted by quantum fluctuations, the investigation of eMChE in the PO state will provide us with insight into effects of quantum chiral spin fluctuations. In Fig. 4**a**, we show $p$-$T$ phase diagram determined from $T$-dependence of resistivity and topological Hall resistivity $\rho_{yx}^{\mathrm{THE}}$ (for the experimental details, see Supplementary Information). The emergence of $\rho_{yx}^{\mathrm{THE}}$ in the PO state is a hallmark of the topological spin correlation endowed with the scalar spin chirality[19]. The $p$-$T$ phase diagram is almost identical to that of bulk sample except for increase of the critical pressure $p_c$ (~ 17 kbar). The increased $p_c$ is probably due to the tensile strain from the Si sample stage, which compensates the effect of applied hydrostatic pressure.

Figure 4**b** shows $T$-dependence of $\rho^{2f}$ in a left-handed MnSi thin plate sample under $B$ = 0.4 T at various pressures, measured with $j = 7.5 \times 10^8\ \mathrm{A/m^2}$. Contour mappings of $\rho^{2f}$ in $T$-$B$ plane are presented at several pressures in Figs. 4**c**-**f**. For $p < p_c \approx$ 17 kbar, a large magnitude of eMChE signal is detected at the periphery just above $T_c$, like the case under the ambient pressure, as seen from Figs. 4**a**–**d** (see also Figs. 3**a** and **b** for $\rho^{2f}$ at $p = 0$ kbar). For $p > p_c \approx 17$ kbar, the eMChE signal shows the maximum magnitude at the lowest measurement temperature within the PO phase, not around the boundaries between the PO and ferromagnetic states nor between the PO and paramagnetic states (Figs. 4**e** and **f**). This feature suggests that the eMChE under $p > p_c$ is induced by the quantum spin fluctuations or dynamics of the PO state, which should possess the chiral nature as well.

In conclusion, thermal and quantum spin fluctuations in the chiral magnet MnSi, which are critically enhanced in association with classical and quantum phase transitions, give rise to asymmetric electron scatterings, leading to the large eMChE. Conversely, the measurement of eMChE can offer a new simple method to study spin fluctuations in chiral spin texture and their related phenomena. In particular, it can be applied for the exploration of dynamic emergent electromagnetic fields evoked by motions of nontrivial chiral spin textures; e.g. skyrmions and emergent magnetic monopoles[20].

## Methods

### Sample preparation

Single crystals of MnSi were synthesized with use of the Czochralski method. Their crystalline chirality was confirmed by using convergent beam electron diffraction (CBED) method (see also Supplementary Information). By focused ion beam (FIB) technique (NB-5000, Hitachi), we cut the thin plates out of those single crystals. Sizes of the thin plates are typically ~10 μm × 20 μm × 500 nm. The thin plates were mounted on a silicon stage and were fixed with the use of FIB-assisted tungsten-deposition. Gold electrodes were patterned by using photolithography and electron beam deposition. We prepared several thin-plate samples to confirm the reproducibility.

**Transport measurements**

Linear longitudinal resistivity and Hall resistivity were measured by using dc-transport option of Physical Property Measurement System (PPMS). Hydrostatic pressures were applied with use of a CuBe clamp cell, and the applied pressures were calibrated with pressure change of superconducting transition temperature of Pb. Second harmonic resistivity was measured by using a Lock-in technique (SR-830, Stanford Research Systems); we input low-frequency (*f*) ac current and measured second harmonic resistivity.

**Acknowledgements** We appreciate T. Ideue, Y.Okamura, and K. Yasuda for useful discussions. This work was supported by JSPS KAKENHI (Grant Nos. 24224009, 24226002, and 15H05456) and CREST, JST.

**Author Contributions** T.Y. fabricated thin plates samples with assistance from K.S. and conducted transport measurements. A.K. grew single crystal. D.M. carried out the CBED method. Y.T. conceived the project. T.Y., N.K., and Y.T. wrote the draft. All authors discussed the results and commented on the manuscript.

**Additional information** Supplementary information is available in the online version of the paper. Reprints and permissions information is available online at www.nature.com/reprints. Correspondence and requests for materials should be addressed to T.Y. and Y.T.

**Competing financial interests** The authors declare no competing financial interests.

**Figure 1 | Crystal and spin structures of MnSi, sample image and experimental configurations for electrical magnetochiral effect. a-c,** Crystal structures of right- and left-handed MnSi viewed from the [111] direction (**a**) and the corresponding spin structures of helical orders (**b**) and skyrmions (**c**). We define the right- and left-handed MnSi as the atomic coordinates ($u$, $u$, $u$), (1/2+$u$, 1/2-$u$, 1/2-$u$), (1/2-$u$, -$u$, 1/2+$u$), (-$u$, 1/2+$u$, 1/2-$u$) with $u_{\mathrm{Mn}}$ = 0.863, $u_{\mathrm{Si}}$ = 0.155 and with $u_{\mathrm{Mn}}$ = 0.137, $u_{\mathrm{Si}}$ = 0.845, respectively. **d,** A scanning electron microscope image of a MnSi thin plate sample: MnSi crystal (green), gold electrodes (yellow), tungsten for fixing the sample (light blue), and silicon stage (grey). **e,** Experimental configurations for measurements of electrical magnetochiral effect and expected dichroic properties of current density. The bold arrows schematically represent paths with the larger current density at a constant electric field along the arrow direction.

**Figure 2 | Electrical magnetochiral effect in MnSi thin plate samples. a**, **b,** Magnetic field dependence of second harmonic resistivity ($\rho^{2f}$) in right-handed (**a**) and left-handed MnSi crystals (**b**). **c**, **d,** Current-density ($j$) dependence of $\rho^{2f}$ (**c**) and angle ($\theta$) dependence of $\rho^{2f}$ (**d**) in left-handed MnSi. Here $\theta$ is the angle between current and the magnetic field as shown in the inset of (**d**). The solid line is fit to $\cos\theta$.

**Figure 3 | Temperature dependence of electrical magnetochiral effect. a,** Contour mapping of second harmonic resistivity ($\rho^{2f}$) in left-handed MnSi in $T$-$B$ plane. The green and blue lines denote the phase boundary enclosing the helical phase and the crossover line between the induced ferromagnetic and paramagnetic phases, respectively. For the $\rho^{2f}$ anomaly around the narrow skyrmion-lattice phase region (denoted by a dotted green line), see Fig. S3 in Supplementary Information. **b,** Temperature dependence of $\rho^{2f}$ at $B$ = 0.4 T.

**Figure 4 │Pressure effect on electrical magnetochiral effect. a,** Pressure ($p$) - Temperature ($T$) phase diagram together with contour mapping of observed topological Hall resistivity $\rho_{yx}^{\mathrm{THE}}$ at 0.4 T (see Fig. S4 in Supplementary Information). **b,** Temperature dependence of second harmonic resistivity ($\rho^{2f}$) for 0.4 T at various pressures in left-handed MnSi. **c-f,** Contour mappings of $\rho^{2f}$ at various pressures in $T$- $B$ plane in left-handed MnSi. The green lines are the phase boundary between the conical and induced ferromagnetic phases determined from magnetoresistivity measurements, and the blue lines are the phase boundary of the partial order phase determined from topological Hall effect measurements.

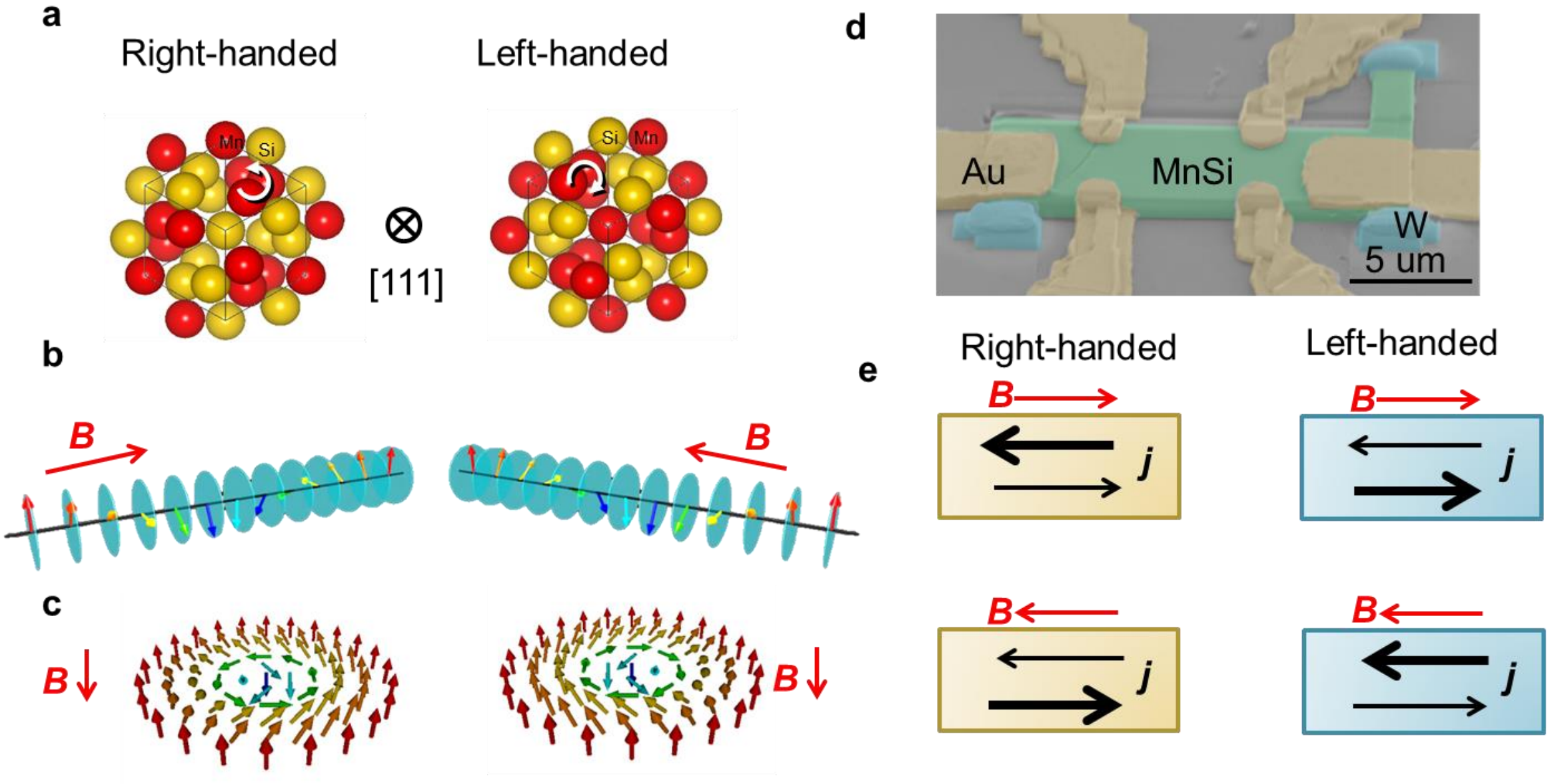

Fig. 1

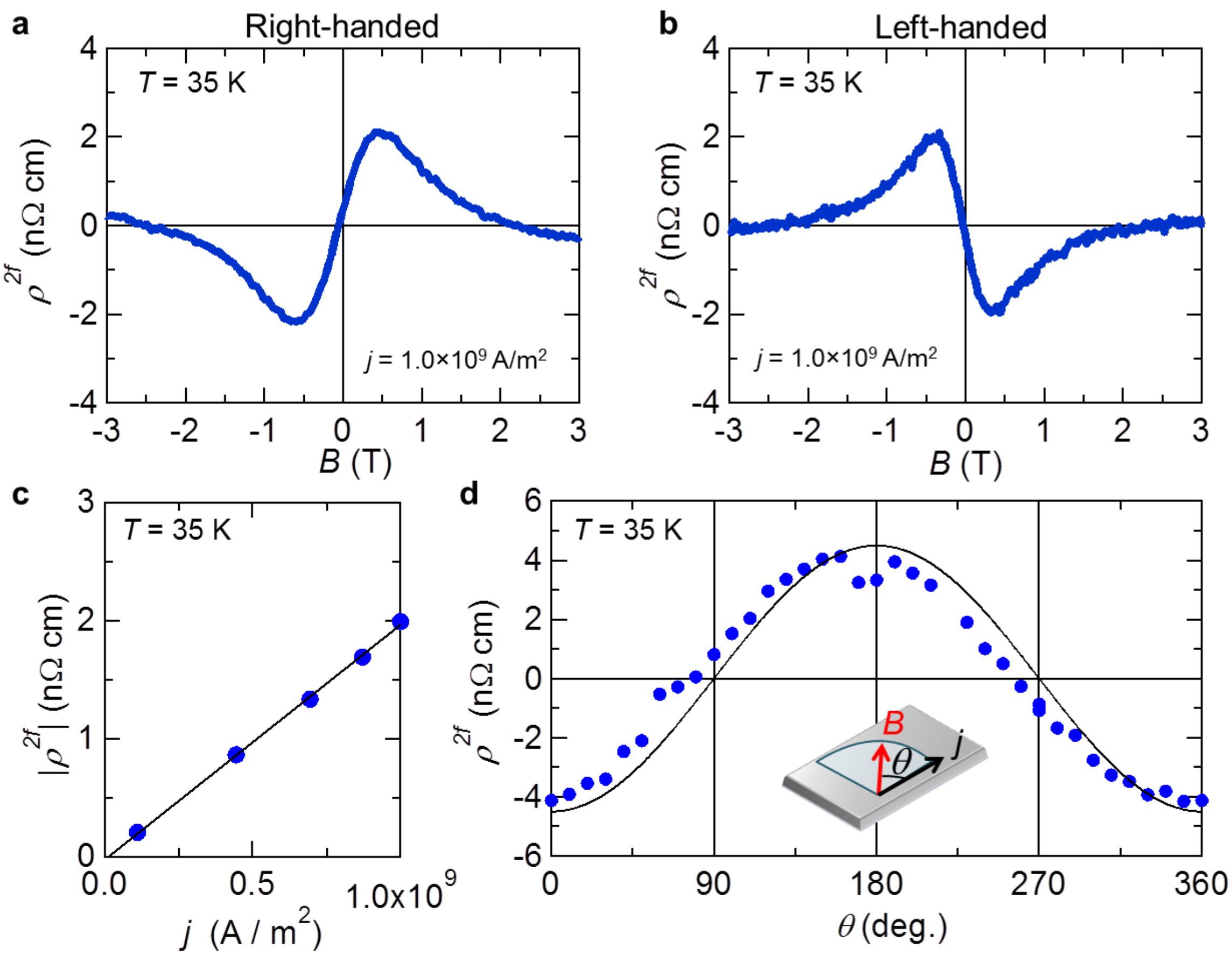
a
Right-handed
T = 35 K
j = 1.0×10⁹ A/m²
ρ^2f (nΩ cm)
B (T)
b
Left-handed
T = 35 K
j = 1.0×10⁹ A/m²
ρ^2f (nΩ cm)
B (T)
c
T = 35 K
|ρ^2f| (nΩ cm)
j (A / m²)
1.0x10⁹
d
T = 35 K
ρ^2f (nΩ cm)
B
θ
j
θ (deg.)

Fig. 2

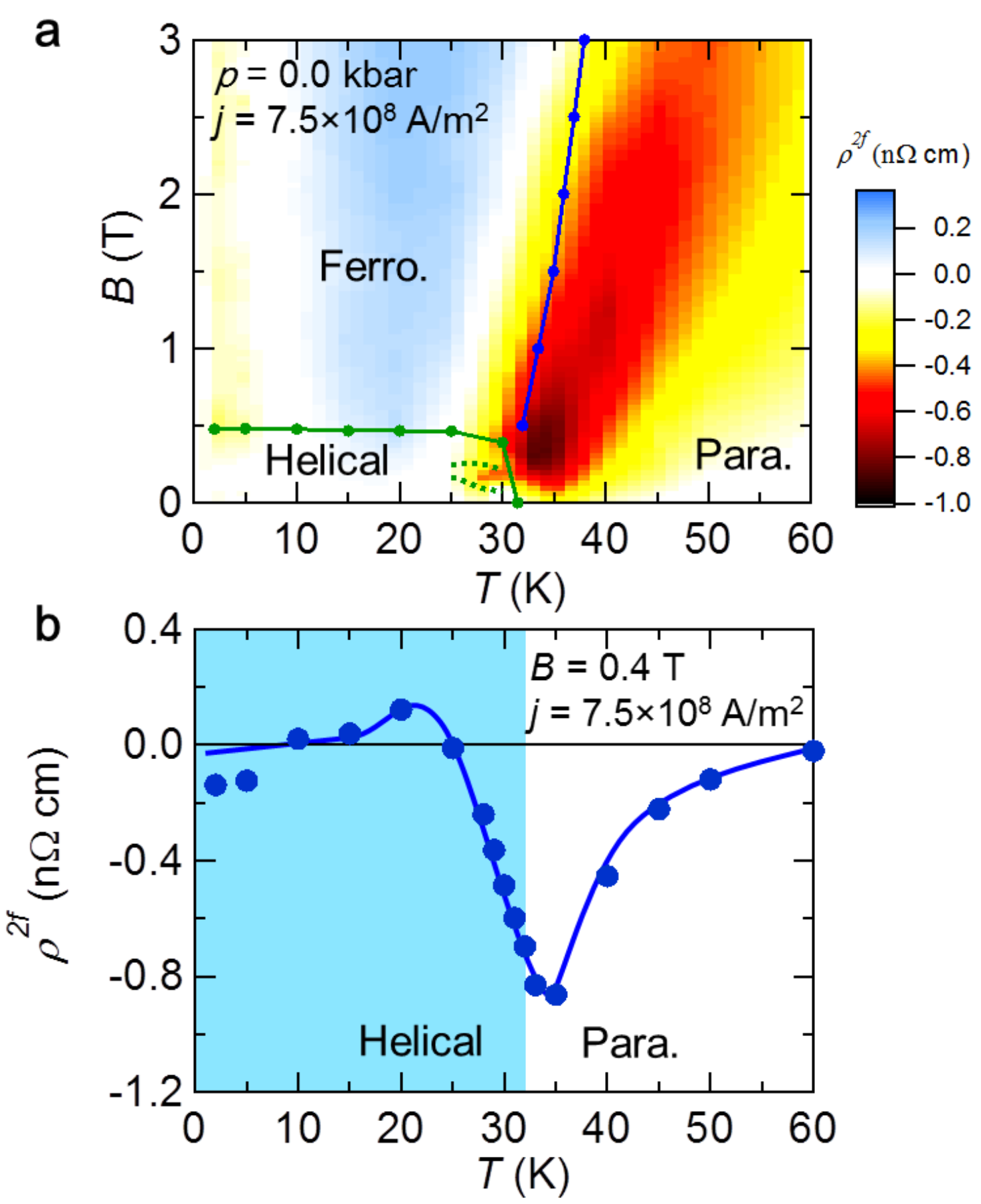
a
p = 0.0 kbar
j = 7.5×10⁸ A/m²
Ferro.
Helical
Para.
B (T)
T (K)
ρ2f (nΩ cm)
b
B = 0.4 T
j = 7.5×10⁸ A/m²
Helical
Para.
ρ2f (nΩ cm)
T (K)

Fig. 3

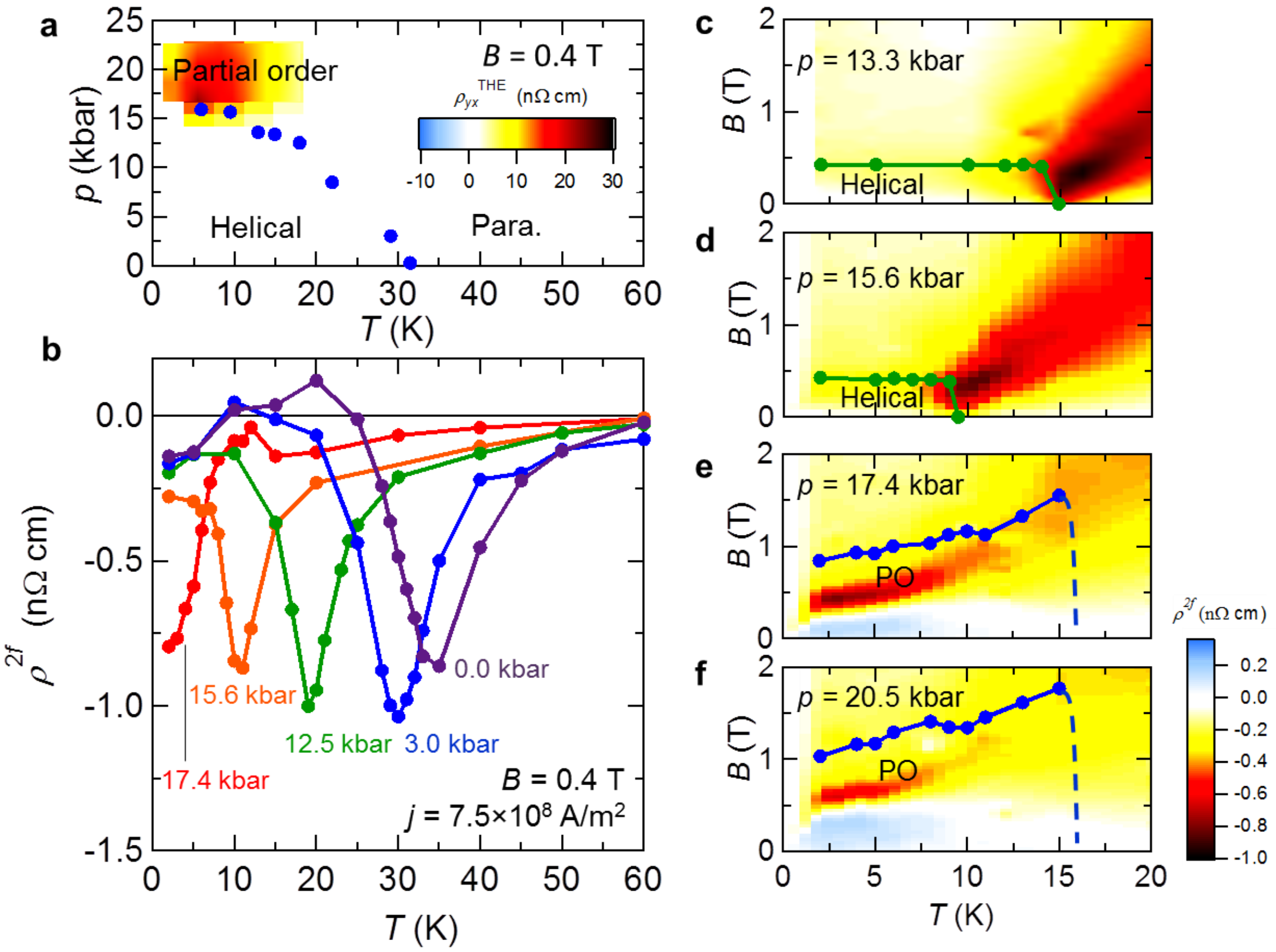
a
B = 0.4 T
Partial order
Helical
Para.
b
B = 0.4 T
j = 7.5×10⁸ A/m²
0.0 kbar
3.0 kbar
12.5 kbar
15.6 kbar
17.4 kbar
c
p = 13.3 kbar
Helical
d
p = 15.6 kbar
Helical
e
p = 17.4 kbar
PO
f
p = 20.5 kbar
PO

Fig. 4

# Supplementary Information for

# Electrical magnetochiral effect induced by chiral spin fluctuations

T. Yokouchi[1†], N. Kanazawa[1], A. Kikkawa[2], D. Morikawa[2], K. Shibata[2], T. Arima[2,3], Y. Taguchi[2], F. Kagawa[2], Y. Tokura[1,2†]

[1]*Department of Applied Physics, The University of Tokyo, Tokyo 113-8656, Japan.*

[2]*RIKEN Center for Emergent Matter Science (CEMS), Wako 351-0198, Japan.*

[3]*Department of Advanced Materials Science, The University of Tokyo, Kashiwa 277-8561, Japan*

† To whom correspondence should be addressed. E-mail: yokouchi@cmr.t.u-tokyo.ac.jp and tokura@riken.jp

**Convergent beam electron diffraction.**

Crystalline chirality of MnSi was confirmed by using convergent beam electron diffraction (CBED) method[1]. Figure S1 shows the observed CBED patterns of the left- and right-handed MnSi crystals in comparison with the simulated ones by using software MBFIT[1].

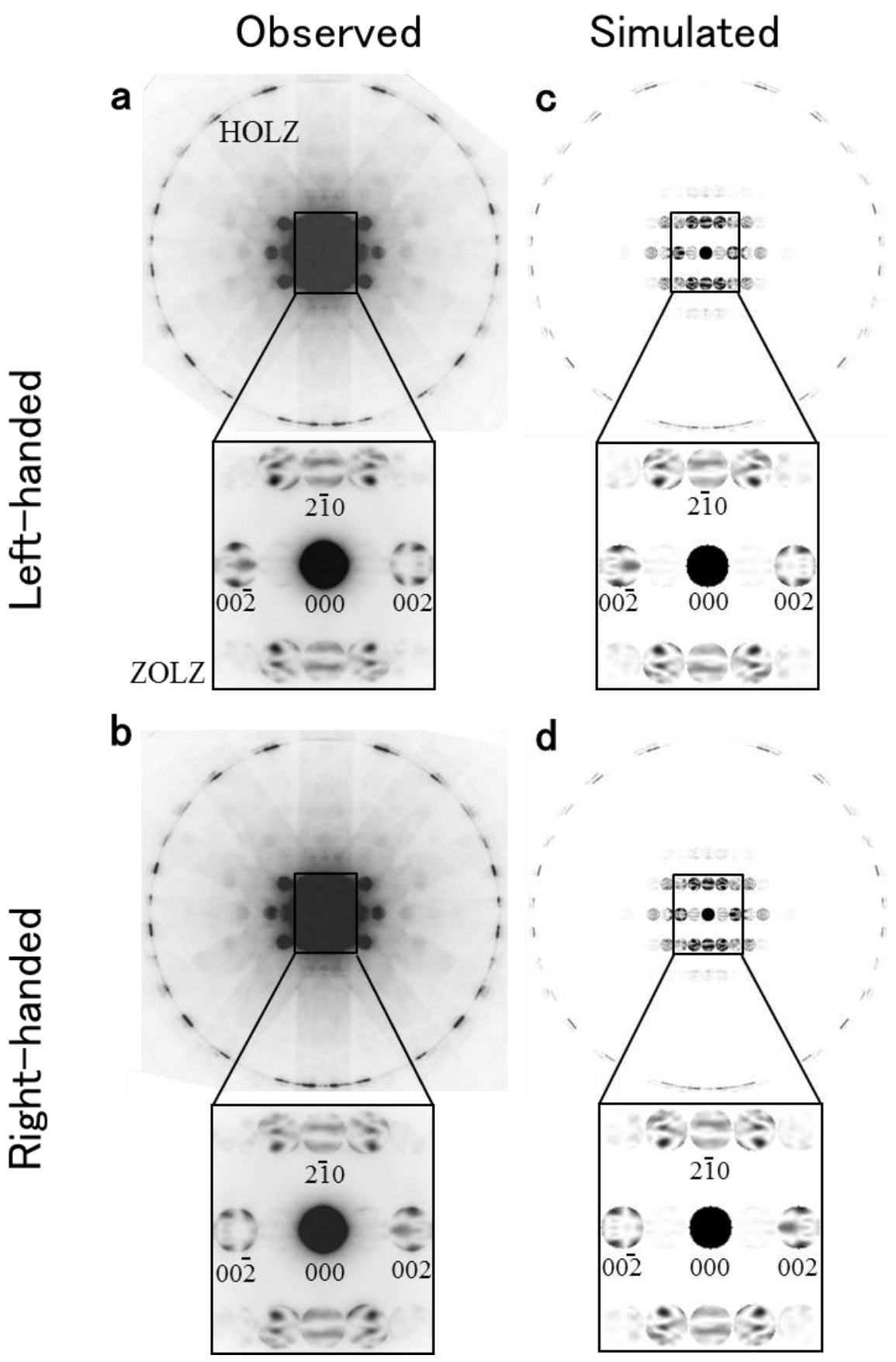

**Figure S1 │Convergent beam electron diffraction patterns (CBED) a**, **b,** Observed CBED patterns taken with [120] incidence from different single crystals. Those patterns are matched to simulated CBED patterns assuming (**c**) the left- and (**d**) right-handed crystal structures, respectively.

**Second harmonic resistivity.**

The second harmonic resistivity is defined as

$$\rho^{2f} = \frac{V^{2f}\,S}{I\,d} = \frac{V^{2f}}{j\,d} \quad \text{(S1).}$$

Here, $V^{2f}$, $I$, $S$, $j$, and $d$ are second harmonic voltage, current, cross-sectional area of the MnSi thin plate, current density, and distance between voltage terminals, respectively. The relation between $\rho^{2f}$ and electrical magnetochiral effect (eMChE) is derived as follows. When we input ac current $\hat{\boldsymbol{j}} = \boldsymbol{j}\sin\omega t$, the electrical magnetochiral voltage $V^{\mathrm{eMChE}}$ is expressed as

$$V^{\mathrm{eMChE}} = \rho\left[\,\hat{\gamma}^{R/L}(\hat{\boldsymbol{j}}\cdot\boldsymbol{B})\right]\hat{j}d = \rho\left[\,\hat{\gamma}^{R/L}(\boldsymbol{j}\cdot\boldsymbol{B})\right]jd\,(\sin\omega t)^2$$
$$= \frac{\rho}{2}\left[\,\hat{\gamma}^{R/L}(\boldsymbol{j}\cdot\boldsymbol{B})\right]jd[1+\cos(2\omega t)]. \quad \text{(S2)}$$

Here the coefficient of $\cos(2\omega t)$ is second harmonic voltage, and therefore we obtain

$$V^{2f} = \frac{\rho}{2}\left[\,\hat{\gamma}^{R/L}(\boldsymbol{j}\cdot\boldsymbol{B})\right]jd. \quad \text{(S3)}$$

From Eqs. (S1) and (S3), $V^{2f}$ is related to eMChE as

$$\rho^{2f} = \frac{V^{2f}}{j\,d} = \frac{\rho}{2}\hat{\gamma}^{R/L}(\boldsymbol{j}\cdot\boldsymbol{B}). \quad \text{(S4)}$$

**Temperature dependence of resistivity in MnSi thin plate samples.**

In Fig. S2**a**, we compare linear longitudinal resistivity ($\rho_{xx}$) of a MnSi thin plate sample with that of bulk single crystal, out of which we sliced the thin plate. Resistivities of two samples show similar temperature (*T*) dependence, indicating minimal damage due to FIB fabrication process. We determined the transition temperature of the helical ordering as the temperature where $\rho_{xx}$-$T$ curve exhibits an inflection. Note that the slight increase of transition temperature in the thin plate sample compared to that of the bulk is due to perhaps uniaxial strains from the silicon sample stage[2].

In Fig. S2**b**, we present *T*-dependence of resistivity of the thin plate sample at various pressures. We also assigned inflection points of $\rho_{xx}$-$T$ curves to the magnetic transition temperatures. The transition temperatures decrease with increasing applied pressure, and above $p$ = 17.4 kbar, the helical transition disappears.

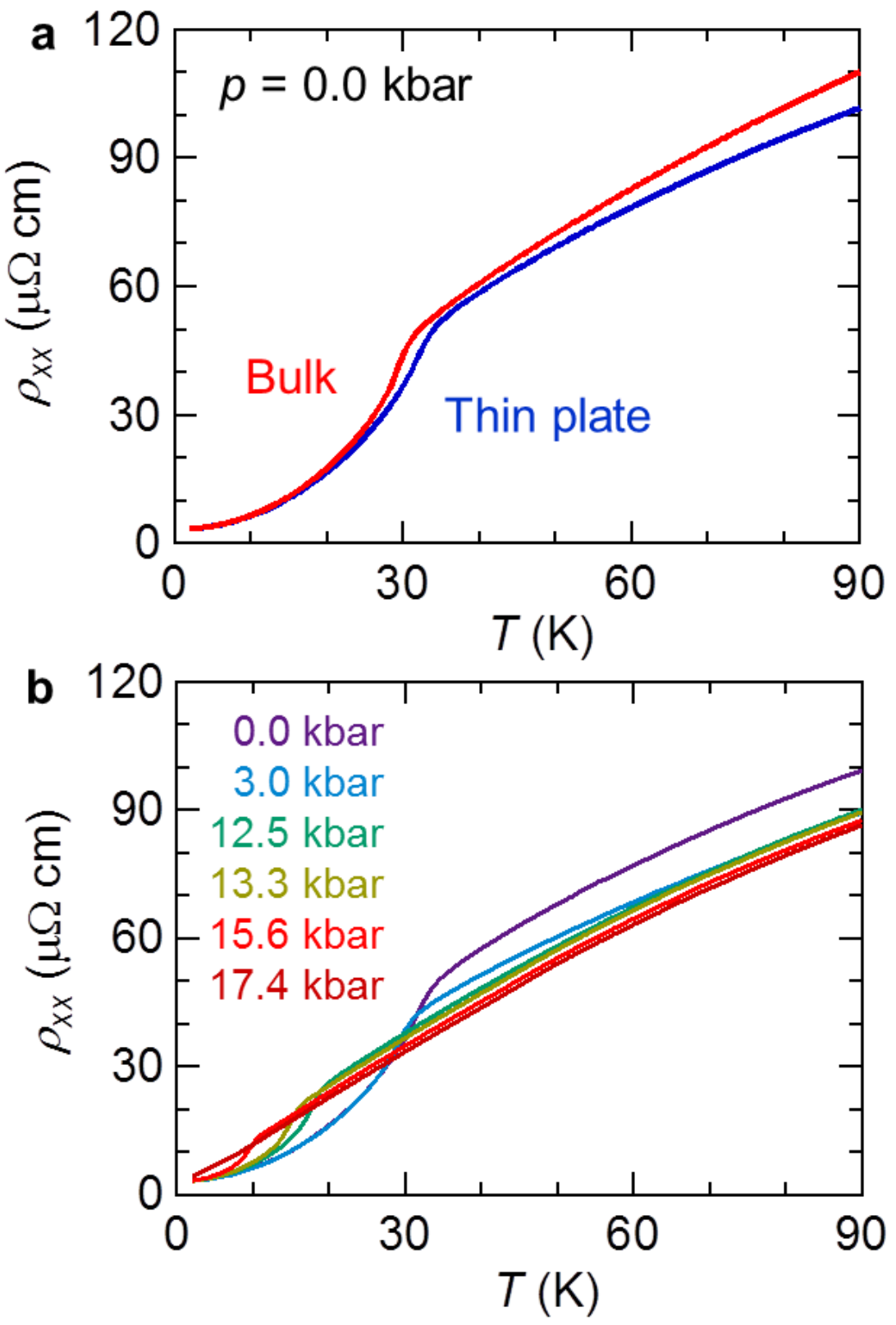

**Figure S2 │Temperature dependence of longitudinal resistivity. a,** Comparison of temperature dependence of resistivity ($\boldsymbol{\rho_{xx}}$) between bulk sample (red line) and thin plate sample of MnSi (blue line). **b,** Resistivity of thin plate sample at various pressures.

**Electrical magnetochiral effect at the phase boundary of the skyrmion-lattice states.**

Electrical magnetochiral effect is also observed at the phase boundary between the conical and skyrmion-lattice states. To precisely estimate the phase boundary, we employed measurements of planar Hall effect (PHE), which was proven in the former study[3] to sensitively detect variations in anisotropic magnetoresistance associated with magnetic transitions, typically showing kinks at phase boundaries between skyrmion-lattice, conical, and induced ferromagnetic states. In Fig. 3S**b**, we present *B*-dependence of planar Hall resistivity ($\rho^{\mathrm{PHE}}$) around $T_{\mathrm{c}}$, marking the phase boundaries as red and green triangles. The magnetic phase diagram for the thin plate sample is also similar to that for bulk crystal, except for slight expansion of the skyrmion-lattice phase region. Incidentally, the stabilization of skyrmion state are attributed to uniaxial strain, which arises from difference in thermal expansion between the MnSi thin plate and the sample stage made of Si (Ref. 4) A contour mapping of $\rho^{2f}$ in a right-handed MnSi sample measured with $j = 1.0 \times 10^9\ \mathrm{A/m^2}$ is presented in Fig. 3S**a**, together with the phase boundaries determined from these $\rho^{\mathrm{PHE}}$-mesurements. This clearly captures the enhanced magnitude of $\rho^{2f}$ at the boundaries of skyrmion-lattice phase, indicating that the asymmetric electron scattering by the chiral spin fluctuations also manifests itself at the phase transition between the skyrmion-lattice and the conical or helical states.

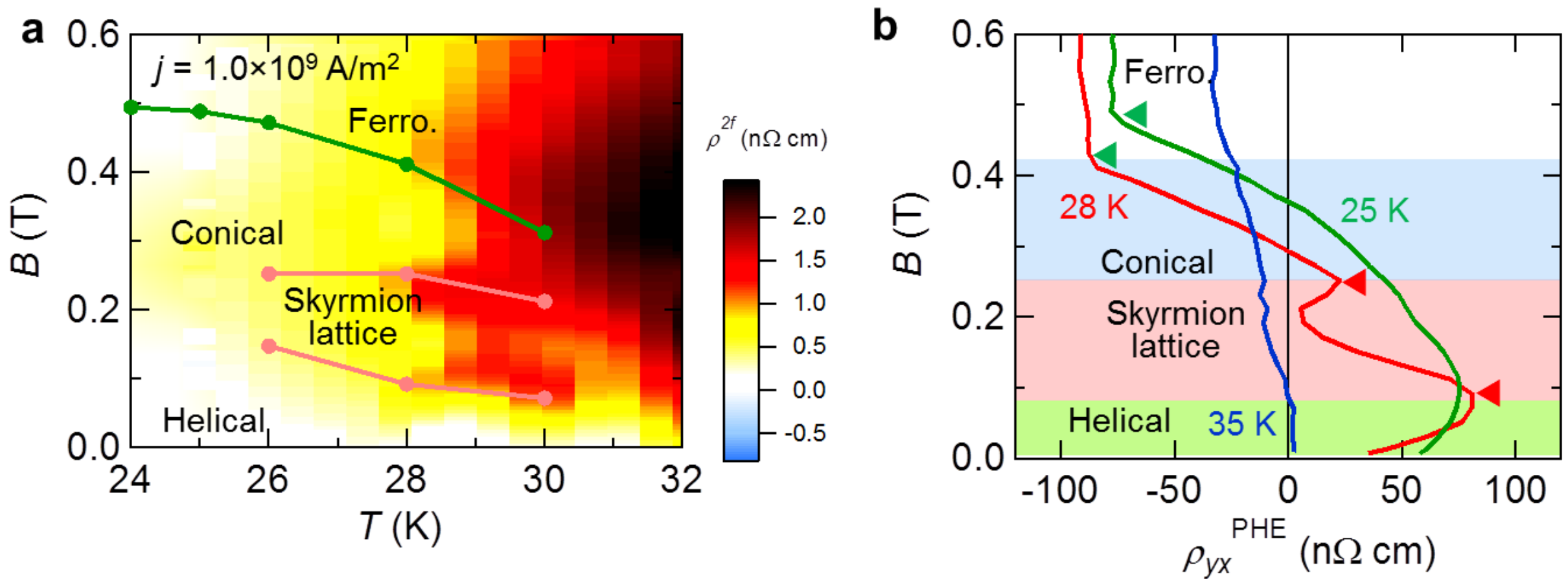

**Figure S3 | Temperature dependence of longitudinal resistivity. a,** Contour mapping of $\rho^{2f}$ of right-handed MnSi around skyrmion phase. **b,** Magnetic field dependence of planar Hall resistivity $\rho^{\mathrm{PHE}}$ at various temperatures. The green and red triangles represent the phase transitions between the ferromagnetic and conical phases and between the conical and skyrmion-lattice phases, respectively.

**Topological Hall effect in MnSi under pressure.**

In conventional magnets, ordinary Hall effect and anomalous Hall effect contribute to Hall resistivity as $\rho_{yx} = R_0B + R_SM$. Here, $B, M, R_0,$ and $R_S$ are magnetic field, magnetization, ordinal Hall coefficient, and anomalous Hall coefficient, respectively. In a noncoplanar spin structure with nonzero topological winding number, such as skyrmions, an additional Hall effect called topological Hall effect shows up[5]. In the case of MnSi, the topological Hall effect is observed not only in the narrow temperature (*T*)-magnetic field (*B*) region of skyrmion phase, but also in a wide *T*-*B* region of the partial order (PO) phase, where applied pressure exceeds the critical pressure for the disappearance of helical order ($p_c$). This indicates the existence of a topological spin structure above $p_c$ (Ref. 6). We measured the Hall resistivity of MnSi thin plates and reproduced the similar signals of topological Hall effect as reported in MnSi bulk sample. In Fig. S4, we show magnetic-field dependence of Hall resistivity in a MnSi thin plate sample at various temperatures and pressures, in accord with the results reported in Ref. 6. Hall resistivity at $p$ = 0 kbar is dominated by sum of ordinary and anomalous Hall signals (Figs. S4**a** – **c**). In contrast, above $p_c$, we found the additional contribution of topological Hall effect as indicated by orange shadows (Figs. S4**d** – **i**). The observation of topological Hall effect identifies the existence of topological spin structure above $p_c$ even in thin plate samples of MnSi.

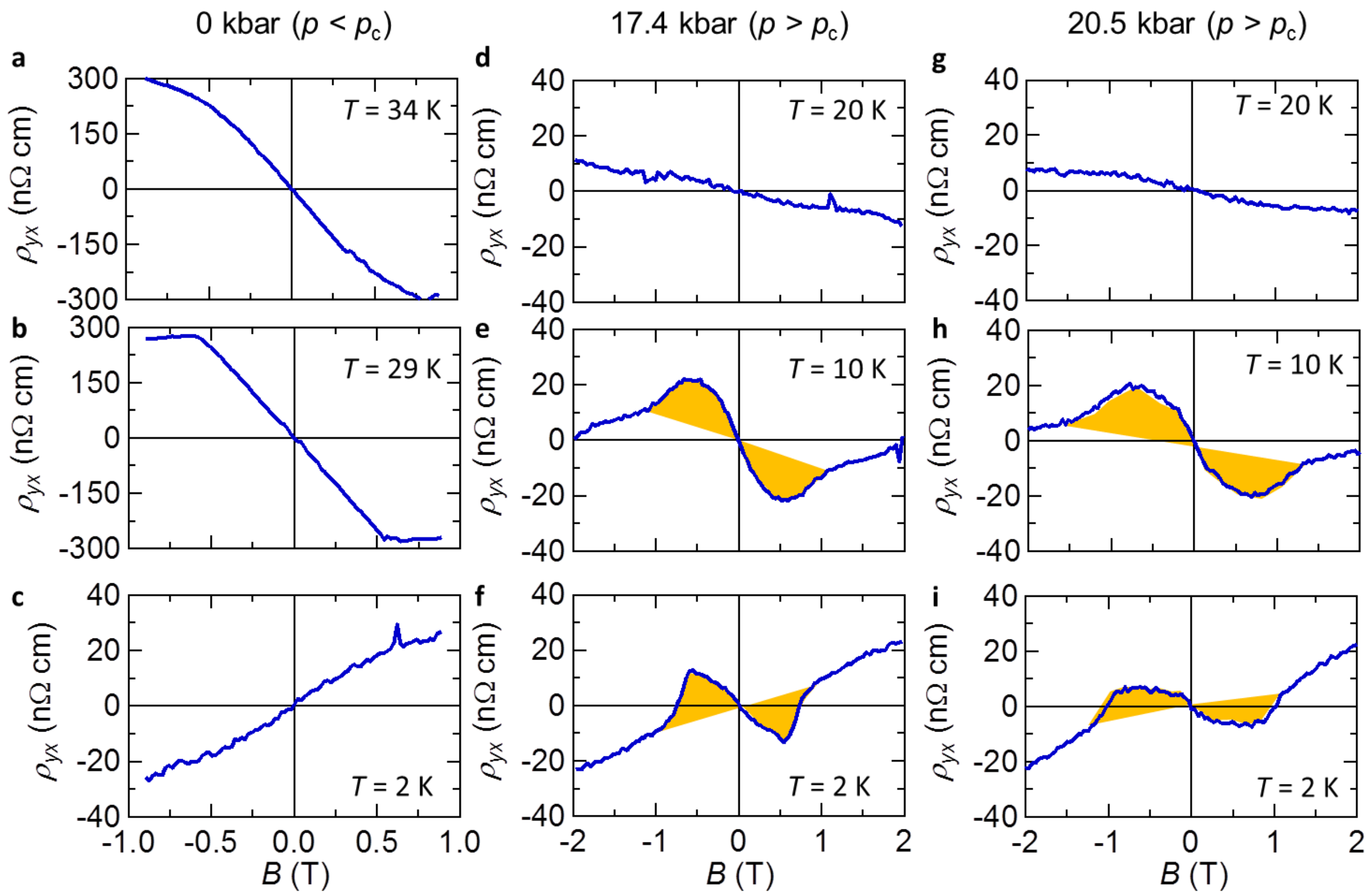

**Figure S4 │Magnetic-field dependence of Hall resistivity at various pressures.**

**Magnetic-field dependence of electrical magnetochiral effect at various temperatures and pressures.**

In Fig. S5, we show magnetic-field dependence of $\rho^{2f}$ in a left-handed MnSi measured with current density $j = 7.5 \times 10^8\ \mathrm{A/m^2}$ at various temperatures and pressures. The helical phases, ferromagnetic phases, and partial magnetic order phases are highlighted by light blue, green, and orange shadows, respectively.

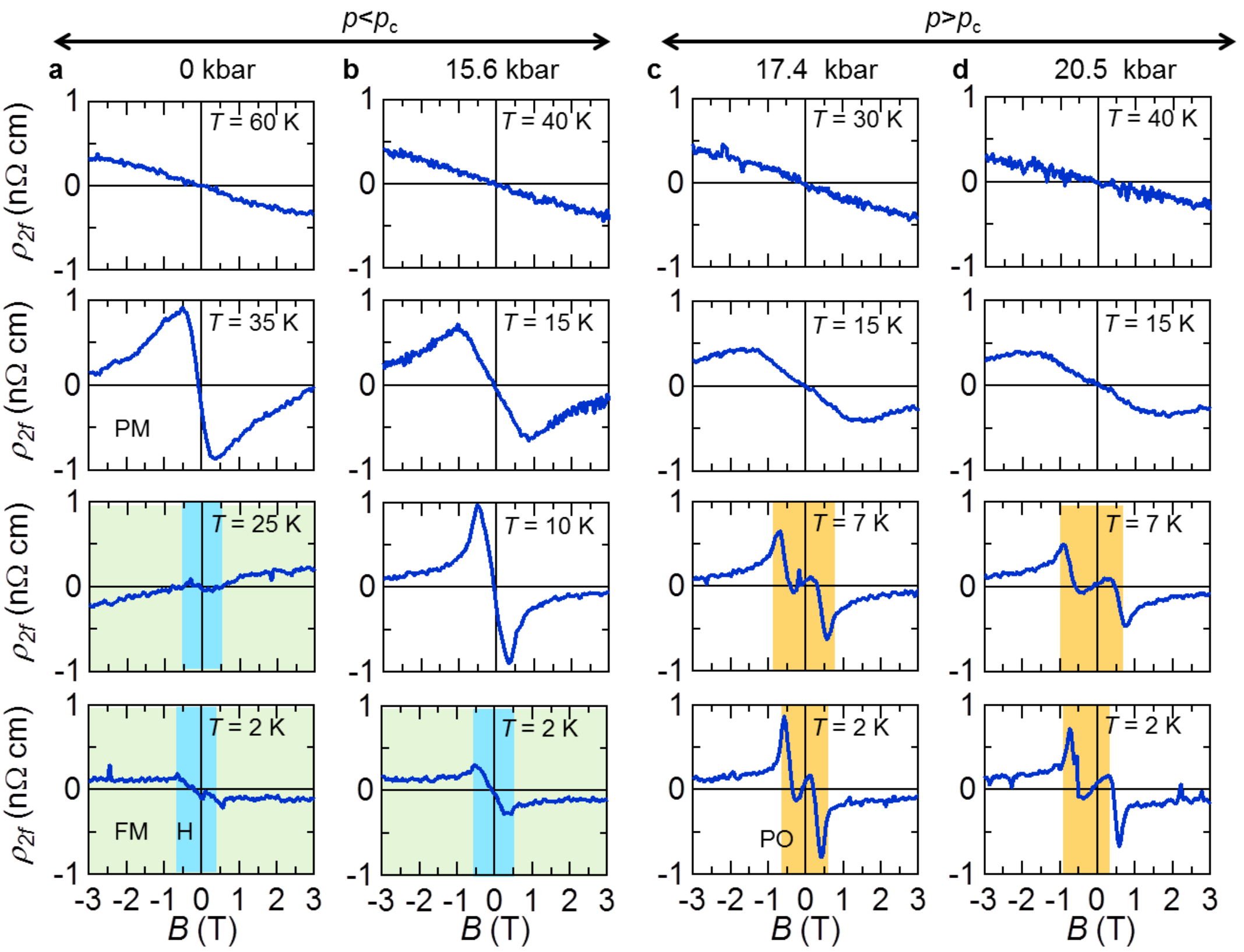

**Figure S5 │Magnetic-field dependence of eMChE at various pressures.**

**References and Notes.**